\documentclass[aps,prl,onecolumn,superscriptaddress,longbibliography]{revtex4-1}
\usepackage{amsmath,amssymb,graphicx,color,wasysym}
\usepackage{textcomp}
\usepackage{ulem}
\usepackage{soul}
\usepackage{color}
\usepackage{xcolor}
\usepackage[toc,page]{appendix}
\usepackage{comment}

\begin{document}

\title{Spatio-Temporal Instabilities of Blood Flow in  a Model Capillary Network}

\author{Mathieu Alonzo}
\affiliation{Universit\'e Grenoble Alpes, CNRS, LIPhy, F-38000 Grenoble, France
}
\affiliation{Universit\'e Grenoble Alpes, CNRS, Grenoble INP, LRP, F-38000 Grenoble, France
}

\author{Nathaniel J. Karst}
\affiliation{Babson College, Wellesley MA, USA}

\author{Thomas Podgorski}
\email{thomas.podgorski@univ-grenoble-alpes.fr}
\affiliation{Universit\'e Grenoble Alpes, CNRS, Grenoble INP, LRP, F-38000 Grenoble, France
}

\author{John B. Geddes}
\affiliation{Olin College of Engineering, Needham MA, USA}

\author{Gwennou Coupier}
\email{gwennou.coupier@univ-grenoble-alpes.fr}
\affiliation{Universit\'e Grenoble Alpes, CNRS, LIPhy, F-38000 Grenoble, France
}

\date{\today}

\begin{abstract}
We present experimental evidence of multiple blood flow configurations in a relatively simple microfluidic network under constant inlet conditions. We provide evidence of multistability and unsteady dynamics and find good agreement with a theoretical {one-dimensional advection} model for blood flow in microvascular networks{ that relies on the widely used laws for rheology and phase separation}. We discuss the ramifications for microfluidic experiments and measurements using blood and implications for \textit{in vivo} microcirculation. Our findings suggest that further modeling in microvascular networks should discard the usual assumption of unique, steady-state flow solutions, with crucial consequences regarding gas, nutrient, and waste transport.
\end{abstract}

\maketitle

\section{Introduction}

Multiphase fluid systems involving non-trivial rheology are often a source of nonlinear phenomena in network flows \cite{ruiz-garcia2021} {as is also the case in models of urban traffic even with idealized configurations \cite{daganzo11}}. At the microscale, the flow of droplets through networks exhibits bistability and oscillations \cite{prakash2007,schindler2008,cybulski15,Cybulski2019}.    In the field of biological fluids, oscillations of microvascular flows have been observed {\textit{in vivo}} \cite{kiani94,Mezentseva16}. Although active regulation mechanisms such as vasomotricity \cite{bagher2011} are involved in local fluctuations of blood flow, the intrinsic stability of a passive microvascular network is an interesting question to address as it might be related to flow configuration in a regulated network.

\begin{figure}
         \includegraphics[width=\linewidth]{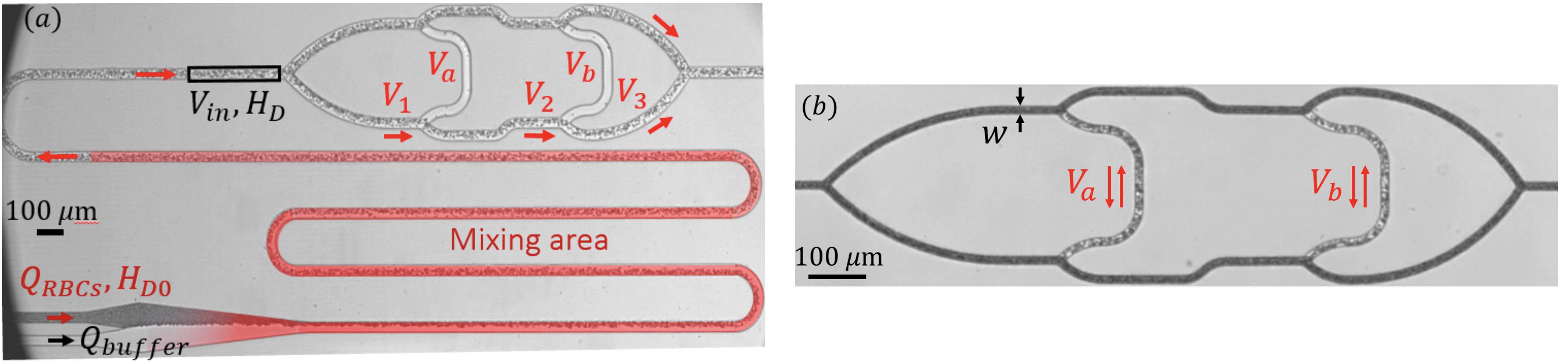}
  \caption{Microfluidic system, 2-bridge model network and notations. (a) The channel inlets (bottom-left corner) allow control of the feed haematocrit
  $H_D$  by mixing of a buffer solution and a concentrated RBC suspension, as well as the inlet average velocity $V_{in}$ that was determined in the depicted zone of interest.
  For large width or low haematocrit (here $w=36~\mu m$, $H_D = 0.2$), the flow splits symmetrically at the first bifurcation
  and no flow is observed in the bridges ($V_a=V_b=0$), similar to an homogeneous Newtonian fluid, while  in (b), for thin enough channels and large haematocrit (here $w=17~\mu m$, $H_D = 0.55$), different flow patterns emerge: a symmetry-breaking takes place with non-zero RBC velocities and concentrations in the bridges. 
\label{fig:scheme}}
\end{figure}

The human blood network architecture is characterized by the superposition of two components: a tree-like network of arterioles and venules that connects to a dense, mesh-like, capillary bed  with diameters ranging from 5 to 100 $\mu$m, \textit{i.e.} roughly 1-10 cell diameters \cite{popel05,lorthois10}.  The multiplicity of paths of various lengths across an organ leads to strong spatial heterogeneities in red blood cell (RBC) and fluid travel times. These various path lengths might lead to heterogeneous oxygenation, and possibly accumulation of various substances involved
in certain diseases \cite{goirand21}.

Blood is a dense suspension of cells, mainly RBCs and a small fraction of white cells and platelets, suspended in plasma. Because of the complex mechanical  properties of blood cells and their interactions, blood dynamics at the cellular scale is inherently fluctuating, either due to RBC shape dynamics \cite{dupire12,mauer18,minetti19}, hydrodynamic interactions and collisions \cite{grandchamp13} or aggregation \cite{brust14,reinhart17}. While these small scale fluctuations may eventually induce perturbations of the flow distribution at the network scale,  it is of fundamental interest to first consider one-dimensional  (1D) models where properties (haematocrit, velocity) have been averaged over the individual vessel's cross section as an initial approach to blood traffic in capillary networks. In this approach, network traffic is modeled by coupling pressure-flow relationships in network branches, which can be described by \textit{e.g.}, empirical laws for the effective viscosity as a function of local haematocrit \cite{pries92} and mass conservation and phase separation laws at bifurcations \cite{dellimore83,fenton85,pries89} that lead to heterogeneity of the haematocrit distribution \cite{doyeux11,balogh2018}.

{The stability of such a strongly coupled network flow problem has been theoretically addressed in the literature \cite{geddes07,davis14,karst15,karst17,davis14_2, benami2022,boissier21,li2023}, revealing multistability and sustained oscillations.  Despite these predictions,  most recent works  on microvascular flows still implicitly postulate the existence of a single steady state \cite{dEsposito18,Schmid19,Mantegazza2020,bernabeu20,goirand21,Enjalbert2024}, without considering the experimental evidence of possible perturbations from the presence of (rare but larger) white blood cells \cite{forouzan12}, local adhesion or aggregation \cite{chang17}, and pathological or impaired cells \cite{Shevkoplyas06}. Notably, no experimental study demonstrating the intrinsic multistability and nonlinear dynamics of blood flow in networks under controlled and steady conditions is reported in the literature.}

 In this work, we present experimental evidence of the emergence of multiple flow patterns in a network and show that their main characteristics are in agreement with a theoretical analysis based on a 1D advection model, without needing to invoke cell-scale fluctuations. To that aim, we analyse the local evolution of velocity and haematocrit distributions in a simple 2-bridge ladder-like symmetric network (Fig. \ref{fig:scheme}), exploring a wide range of inlet haematocrit and various channel widths.

 \section{Experimental method}
 We detail here the experimental methods, leaving the details to the appendix \ref{sec:appendix-exp}. The microfluidic chip was fabricated using classic soft-lithography techniques 
 by casting polydimethylsiloxane on a SU-8 master produced by direct laser lithography (\textit{Klo\' e, Dilase 250}), and then bonded to a plasma-treated glass slide. All channels in a given 2-bridge network have equal length $L=0.5$~mm, equal height $h=30~\mu$m, and equal width $w \in [11,36]~\mu$m. 
 Blood samples were collected by \textit{Etablissement Fran\c{c}ais du Sang}
 in citrate tubes to prevent coagulation. RBCs were separated by centrifugation and re-suspended in a buffer solution made of 68.5\% of nominal phosphate-buffered saline (PBS) mixed with 31.5\% Optiprep (v/v) and 2 g/L  Bovine Serum Albumine. This follows a recently standardized protocol that ensures RBC conservation while also preventing sedimentation \cite{merlo23}. 
 
The inlet haematocrit $H_D$ was continuously adjusted by varying
 the relative flow rates of two separate inlets, achieving a mixing of concentrated RBCs with discharge (reservoir) haematrocit $H_{D0}\simeq 60-70 \%$ and the buffer solution (Fig. \ref{fig:scheme}(a)).  A critical aspect of the experimental procedure is the ability to quickly rinse the system with buffer solution
 immediately after an acquisition, at constant wall shear stress, thus confirming the absence of clogging and validating that observed phenomena are intrinsic to the nonlinear dynamics arising from rheology/phase separation coupling (see Movie in \cite{suppl}).

\begin{figure}
\includegraphics[width= \linewidth]{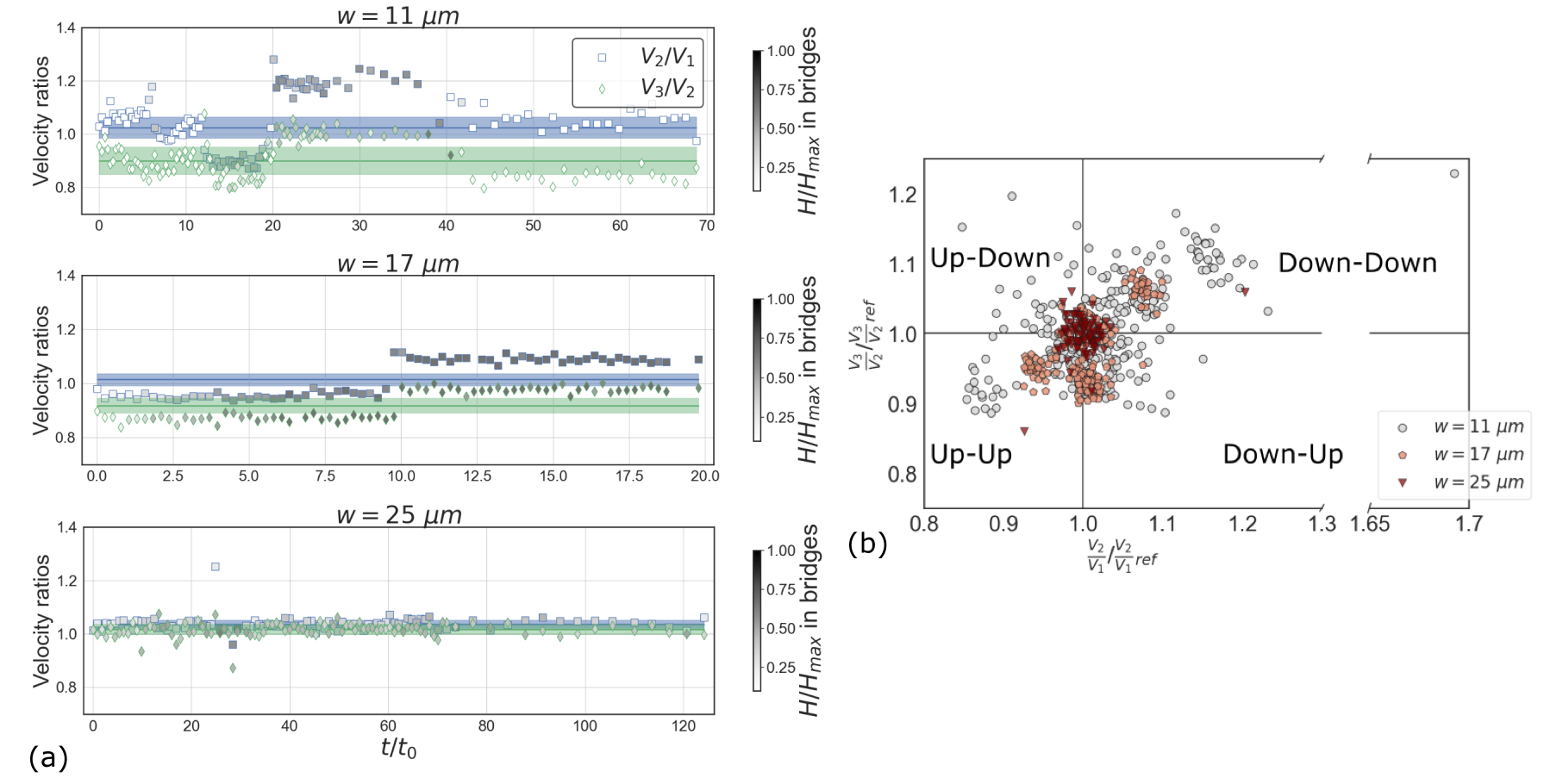}
\caption{ (a) Experimental velocity ratios $V_2/V_1$ and $V_3/V_2$ as a function of normalized time, for three channel widths. The inlet haematocrit is constant in time ($H_D\simeq 0.65 \pm 1\%$, see Fig. \ref{fig:V2V1_ft_complement}(a) in appendix \ref{sec:appendix-ressup}) and is chosen large enough to highlight clear behaviours out of the measurement noise. Concentrations in each bridge, normalized by the maximum concentration, are shown in gray scale. Colored bands identify the reference state, accounting for measurement fluctuations. Other acquisitions for the $w=11 \,\mu$m network are shown in Fig. \ref{fig:V2V1_ft_complement}(b) and exhibit similar non-trivial dynamics. (b) The same data complemented by two other times series for $w=11\,\mu$m and one time series for $w=17\,\mu$m, plotted in the velocity ratio space. ``Up" or ``Down" couples in each quadrant refer to the direction of flow in the bridges.
  \label{fig:V_ratios_function_time_and_concs_ALL}}
\end{figure}

The discharge haematocrit $H_D$ in the inlet branch, \textit{i.e.}, the ratio of RBC flow to total flow, was determined {by measuring the tube haematocrit} by light absorption {\cite{grandchamp13,Merlo2022}, which is then }corrected for the F\r{a}hr\ae{}us effect \cite{Fahraeus1929}. A PIV-like method based on the intensity patterns of the RBCs was developed to measure the maximum (central) velocity in the inlet ($V_{\textrm{in}}$, of the order of the mm/s) and in the main branches ($V_1$ to $V_3$, see Fig. \ref{fig:scheme}). In the following, time is rescaled by the bulk transit time 
$t_0=8 L/\langle V\rangle$, 
where $\langle V \rangle$ is the mean inlet velocity, approximated to $V_{in}/2$ in the experiments. 
Deviations  of the velocity ratios $V_2/V_1$ and $V_3/V_2$ from 1 indicate the existence of flux in the bridges; this may happen even if no cells enter the bridge, since a low flow in the bridge will be fed only by the cell-depleted fluid flowing near the wall of the upstream branch \cite{fedosov10,losserand19}. 

\section{Multiplicity of flow solutions}

The flow solution for a homogeneous, Newtonian fluid flowing in our model symmetric network, hereafter called the reference state, is trivial with $V_a=V_b=0$. 
Here, when channels are thin enough and haematocrit is above a threshold, the observation of the network over long times with fixed inlet conditions reveals a rich dynamics featuring successions of apparently stable asymmetric states and quick spontaneous transitions involving flow reversal in the bridges (Fig. \ref{fig:V_ratios_function_time_and_concs_ALL}(a)). 
Noteworthy, some states may persist over extremely long times compared to the typical transit time $t_0$. For the thinner channel, Fig. \ref{fig:V_ratios_function_time_and_concs_ALL}(b) shows that velocity ratios  fluctuate between $0.8$ and $1.2$, indicating non-negligible flux in both bridges.  Because of slight imperfections in the network resulting from the manufacturing process, the reference state does not correspond to velocity ratios strictly equal to 1. Once identified (see appendix \ref{sec:appendix-exp}) this reference state is used to normalize results for comparison between different networks. 

\begin{figure}
    
     \includegraphics[width=0.7\linewidth]{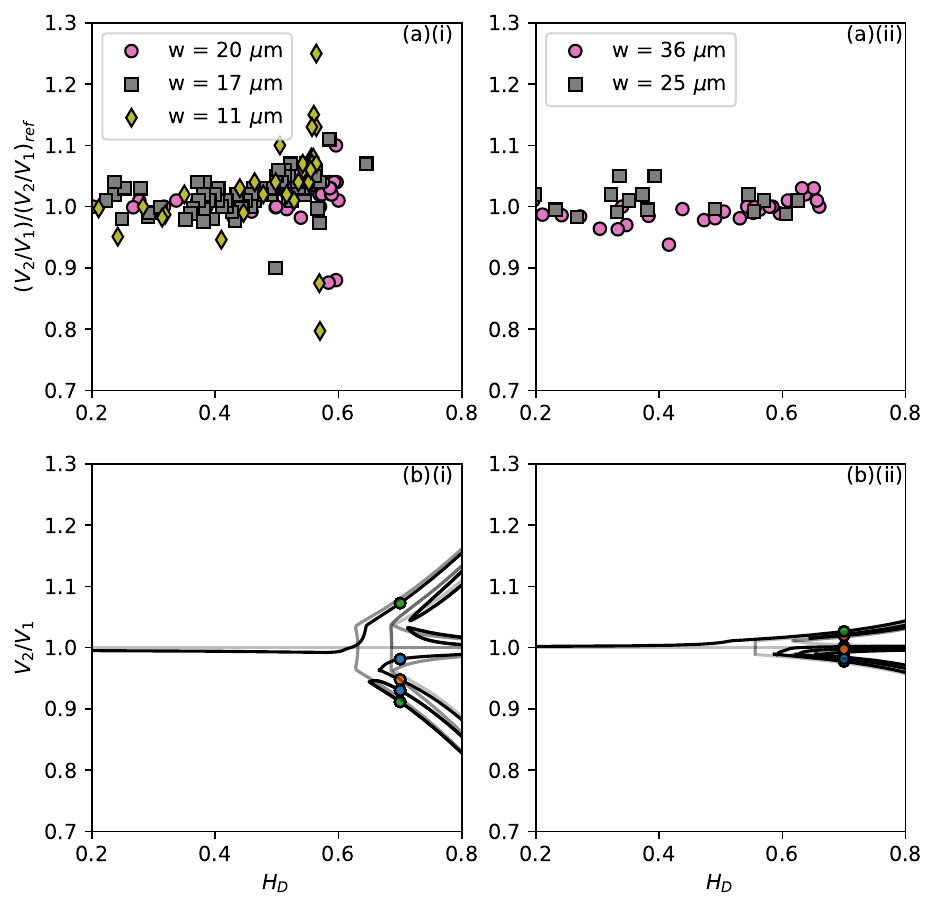}

  \caption{(a) Experimental velocity ratio $V_2/V_1$ normalized by that in the reference state, versus inlet haematocrit $H_D$, for different channel widths $w$: (i) thin channels, $w\le 20\,\mu$m  and (ii) thicker channels, $w \ge 25 \,\mu$m. The velocities are measured a few $t_0$ after the flow is established. (b) Theoretical bifurcation diagram for (i) a geometrically symmetric (light) and perturbed network (dark) with $w=11 \, \mu$m and (ii) a geometrically symmetric (light) and perturbed network (dark) with $w=36~\mu $m. Colored marks correspond to the equilibria supported at $H_D = 0.7 $, also reported in Fig. \ref{fig:stab-sim}(a).\label{fig:V2V1_f_H} }
\end{figure}

Fig. \ref{fig:V2V1_f_H}(a) reports such a comparison while also considering the influence of the feed haematocrit $H_D$. It displays a plot of the velocity ratio $V_2/V_1$ as a function of $H_D$ for a single acquisition beginning about $30$ s ($\,\sim\,10\,t_0$) after blood injection in the network.  For narrow enough channels  ($w\le20~\mu$m), Fig. \ref{fig:V2V1_f_H}(a-i) features two regions: a plateau for $H_D \lesssim 0.45$  {(tube haematocrit $H_T \lesssim 0.33$ for $w=11\,\mu$m)}
followed by
an apparently multi-valued scatter, which is absent for larger channels (Fig. \ref{fig:V2V1_f_H}(a-ii)). These results demonstrate 
that the existence of multiple nontrivial states 
is triggered by crossing thresholds in channel width and inlet haematocrit.  

Due to the fixed measurement time, these states can be either transient or converged. Because of the long residence times in  different states, an extensive and statistically complete study of the nature of these states  would require long-lasting experiments (of the order of an hour) that would most probably be interrupted by unavoidable clogging events in such \textit{in vitro} conditions. This experimental constraint makes a thorough experimental study beyond the scope of this article. We show below that the experimental evidence of a rich dynamics characterized by transitions between multiple states is well supported by theory.

\section{Mathematical model}

We base our theoretical analysis on a  previously established 
one-dimensional advection model for the transport of haematocrit through the network \cite{Pop:2007}. The complex nature of blood flow reveals this model to be fundamentally nonlinear: red cell separation at diverging junctions depends on the flow, diameter, and haematocrit of surrounding vessels; and blood viscosity depends on red cell concentration. Resistance to flow, which determines the advection velocity in every vessel, therefore depends on the haematocrit distribution throughout the entire network. As a result, there may exist multiple equilibrium solutions for flow and haematocrit in the network, and these solutions may themselves lose stability to periodic oscillations \cite{ben-ami:2022}.

More precisely, if we normalize by the inlet flow, the network under consideration has 9 distinct flows $Q_i$, which are constrained by conservation of mass at each of the 6 nodes in the network. We close the system by identifying three Kirchoff equations $\sum_i Q_i R_i = 0$, where $R_i$ is the  resistance in vessel $i$ and is given by Poiseuille's law,
\begin{eqnarray}
    R_i = \frac{128 L}{\pi w^4} \eta_i,
\end{eqnarray}
where $L$ and $w$ are the length and diameter of the vessel, respectively, and $\eta_i$ is the  viscosity. We use a well-known empirical model for \textit{in vitro} viscosity as a function of both haematocrit and the vessel's geometric properties~\cite{pries89,pries92}:

\begin{align}
\eta_{\mbox{vitro}} &= 1 + (\eta^* - 1){(1-H_D)^C - 1 \over (1 - 0.45)^C - 1} \\
\eta^* &= 220 e^{-1.3w} + 3.2 - 2.44e^{-0.06w^{0.645}} \\
C &= 0.8 + e^{-0.075w}(y-1) + y \\
y &= {1 \over 1 + 10^{-11}w^{12}},
\end{align}
where $w$ is again the vessel diameter measured in microns, and $H_D$ is the discharge haematocrit. The effective viscosity of the fluid is then given by $\eta = \eta_{\mbox{vitro}} \eta_0$, where $\eta_0$ is the viscosity of plasma (whose value does not modify the nature of flow solutions).

The discharge haematocrit in vessel $i$ is determined by either conservation of RBC flow if vessel $i$ is the outflow of a converging node,
\begin{eqnarray}
    Q_i H_i = H_j Q_j + H_k Q_k,
\end{eqnarray}
where vessels $j$ and $k$ are the feed vessels, or via plasma skimming if vessel $i$ is the outflow of a diverging node,
\begin{eqnarray}
    Q_i H_i = Q_j H_j f(\cdot),
\end{eqnarray}
where vessel $j$ is the feed vessel. In general, the plasma skimming function $f$ depends on the flow rate in the daughter vessel relative to the flow in the feed vessel $j$, with lower flows recruiting significantly lower proportions of the total red blood cell balance. Here, we use the empirical law of Pries et al. \cite{pries89} that also accounts for differences in red cell distribution as it relates to vessel diameters and feed haematocrit:
\begin{align}
f = {1 \over Q} \begin{cases}
0, \hfill Q < Q_0 \\
\left[1 + \exp\left(A - B\log\left(Q-Q_0 \over 1-Q - Q_0\right)\right)\right]^{-1}, \hfill ~~Q_0 \leq Q \leq 1 - Q_0 \\
1, \hfill Q > 1 - Q_0.
\end{cases} \label{eqn:skim}
\end{align}
where $Q = |Q_i / Q_j|$ is the fractional flow entering daughter branch $i$ instead of the alternative daughter branch $i'$, and the empirical  constants have been constructed to capture experimental data:
\begin{align}
A &= {6.96 \over w_i} \ln \left({w_i \over w_{i'}}\right) \\
B &= 1 + 6.98 \left({1 - H_j \over w_j}\right)\\
Q_0 &= {0.4 \over w_j}.
\end{align}

In our situation where all channels have equal widths, $A=0$. Each resistance featured in our three Kirchoff equations can be expressed solely as a function of the three free flows, and so we have closed the system of nonlinear equations. The determination of the equilibrium flows and haematocrits in networks has been well-detailed elsewhere \cite{karst15}, and we use numerical continuation methods to track the equilibrium solutions as we change system parameters. The fully symmetric network exhibits non-generic bifurcations \cite{kuznetsov}, and so we also consider geometrically perturbed systems that more closely match experiments; here we introduce random deviations of less than 1\% to all vessel diameters.

\begin{figure}
\includegraphics[width=0.7\textwidth]{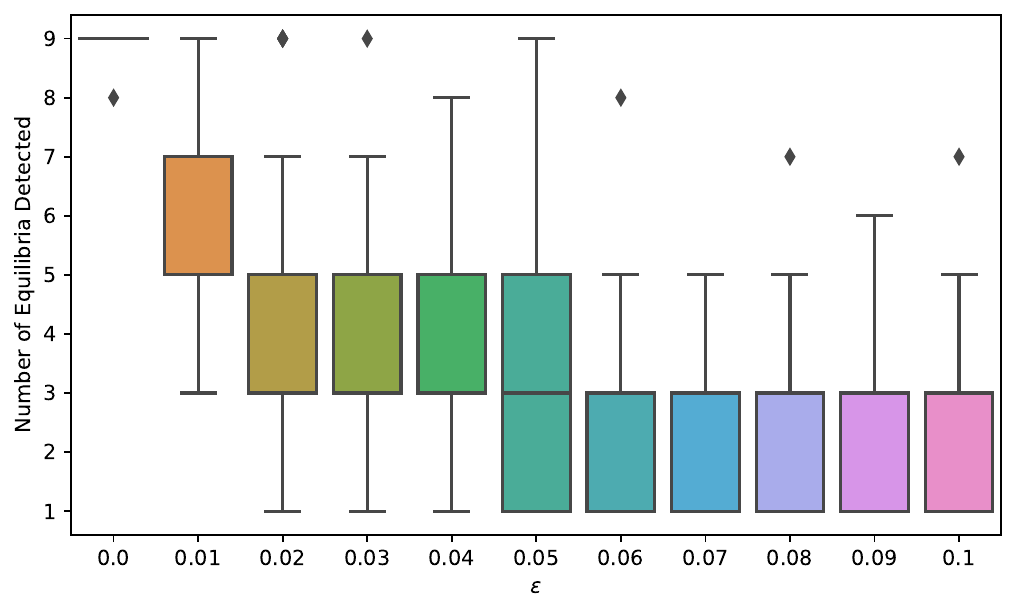}
\caption{Box plots of number of equilibria in geometrically perturbed networks of reference diameter 11 $\mu$m, with diamonds indicating outlier observations. We generated $100$ random networks with vessel diameters perturbed by $\epsilon$ up to $10 \%$. For each, we sample $200$ different initial conditions and iterate until convergence. An equilibrium is saved only if the solution quality is high and the equilibria are well differentiated from one another.}
\label{fig:statistical}
\end{figure}

We show in Fig. \ref{fig:V2V1_f_H}(b) the equilibrium solutions for two 2-bridge networks of channel diameters $w=11$ and 36 $\mu$m respectively, for different inlet haematocrit values. Notice that for the geometrically perturbed system the solution curves become asymmetric but otherwise retain the same features as the fully symmetric case.  We have confirmed the generality of these predictions by conducting a statistical study with perturbed nominal resistances via the diameters: $\tilde{w} = 11 + u\epsilon$, where $u$ is uniformly sampled from the unit interval and $\epsilon$ is the strength of the perturbation. For each value of $\epsilon$, we generate 100 such nominal resistance profiles. For each nominal resistance profile, we generate a collection of equilibria by generating 200 haematocrit profiles (sampled uniformly randomly from the unit interval for each vessel) and use each of these distributions as the initial condition for a standard nonlinear root finder for the equilibrium equations $F(\mathbf{h}) = \mathbf{0}$, where $\mathbf{F}$ is the equilibrium relation and $\mathbf{h}$ is the vector of haematocrits in each vessel. We keep a candidate $\mathbf{h}^*$ only if the solution quality is high ($\|F(\mathbf{h^*})\|_2 \leq 1\times 10^{-6}$) and the candidate is well differentiated from all others ($\|\mathbf{h^*} - \mathbf{h^*}'\|_2 \geq 1 \times 10^{-3}$). In Fig. \ref{fig:statistical} we show the results of this analysis. We see that multiple equilibria are supported across all perturbation strengths, but that the number of total equilibria that are supported decreases as the perturbation strength is increased. 

\begin{figure}
\includegraphics[width=\textwidth]{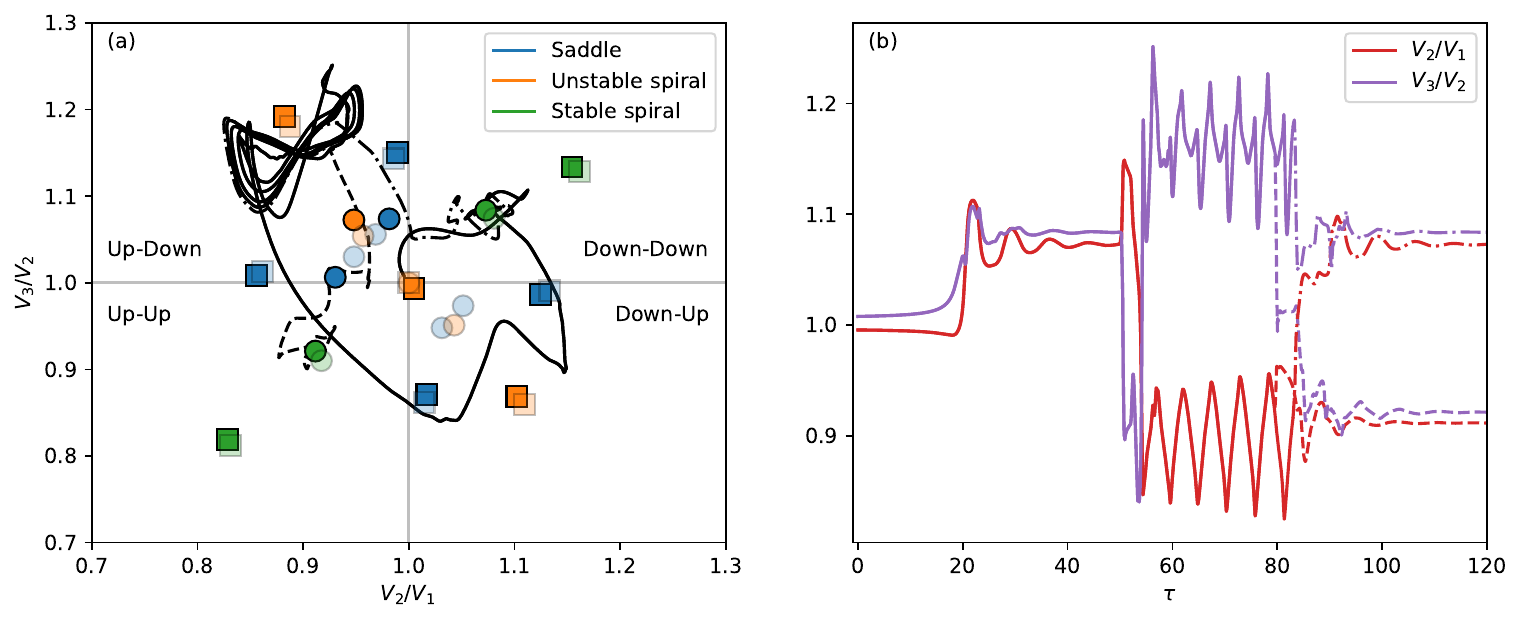}

\caption{(a): Stability of the equilibria supported by the symmetric networks (semitransparent marks) and perturbed networks (opaque marks) for $H_D = 0.7$ (circular marks) and $H_D = 0.8$ (square marks) with $w = 11~\mu m$ in all cases.  The traces represent the trajectories from the time domain simulations in the geometrically perturbed network shown in (b): velocity ratios are plotted as a function of normalized time $\tau=t/t_0$. After ramping up, the inlet haematocrit is $H_D=0.7$ for $20 < \tau < 50$, then $H_D = 0.8$ for 29 transits (dashed trace) or 33 transits (dot-dashed trace), before returning to $H_D=0.7$. }

\label{fig:stab-sim}
\end{figure}

The bifurcation diagram of Fig. \ref{fig:V2V1_f_H}(b) confirms the presence of a threshold in both channel diameter and inlet haematocrit for the existence of non-trivial equilibrium solutions in the network. Note that the threshold haematocrit at which multiple solutions appear in the simulation results are noticeably higher than in experiments ($H_D \simeq 0.62$ vs. $H_D \simeq 0.45$ for $w=11$~$\mu$m). This may be related to the choice of the viscosity \cite{Pries94} and plasma skimming models established from rather scattered data obtained in cylindrical channels, that should not be expected to perfectly describe the behavior in channels with rectangular cross-sections. At $H_D = 0.7$ and $w=11\,\mu$m, there are five distinct equilibria supported by the perturbed network, and each of these may be characterized by the flow direction in the bridges --- these are shown in the $(V_2/V_1, V_3/V_2)$ space in Fig. \ref{fig:stab-sim}(a). We can characterize the system by the flow direction in the bridges, e.g. Up-Up (UU), Up-Down (UD), \textit{etc.}

\begin{figure}
    \includegraphics[width=0.7\textwidth]{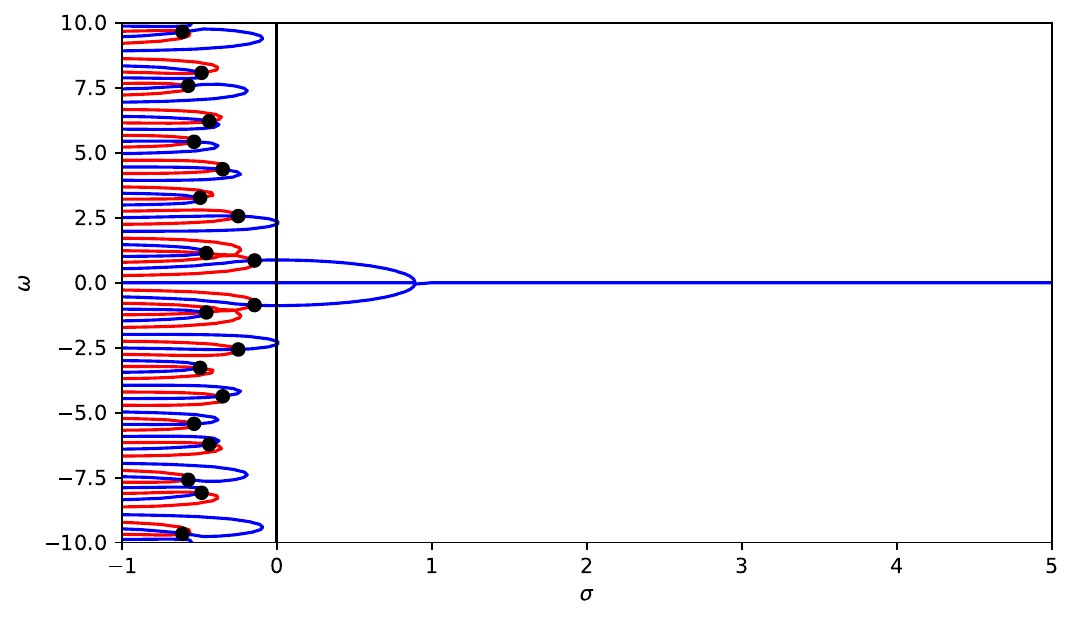}
    \caption{Zero contours of the real (red traces) and imaginary (blue traces) parts of Eq. \ref{eqn:characteristic} evaluated at $(V_2/V_1, V_3/V_2) \approx (1.072,1.083)$ in the perturbed system of Fig. \ref{fig:V2V1_f_H}(b-i). Black marks indicate eigenvalues of the system.} 
    \label{fig:eigencontours}
\end{figure}

Stability of the equilibrium flow distribution is determined by linearizing the dynamic advection model for RBC transport through the network,
\begin{eqnarray}
\frac{\partial H_i}{\partial t} + V_i \frac{\partial H_i}{\partial x_i} = 0,
\end{eqnarray}
where $V_i$ is the advection speed through vessel $i$. The result is a transcendental characteristic equation,
\begin{eqnarray}
1 = \sum_j \frac{c_j}{\tau_j} (1 - e^{-\lambda \tau_j}),
\label{eqn:characteristic}
\end{eqnarray}
where $\tau_j$ is the transit time of every partial flow pathway through the network and $c_j$ is determined by linearizing the effects of plasma skimming and viscosity on the network response \cite{davis14}. The eigenvalues $\lambda = \sigma + i \omega$ are complex and are challenging to compute in general. For illustration purposes, we show an example in Fig. \ref{fig:eigencontours} of the real and imaginary contours of Eq. \ref{eqn:characteristic} evaluated at $(V_2/V_1, V_3/V_2) \approx (1.068,1.090)$ in the perturbed system, \textit{i.e.}, a green mark in Fig. \ref{fig:stab-sim}. Eigenvalues correspond to the intersection points of the two curve sets, and so we conclude that this equilibrium is a stable spiral, as its eigenvalues with largest real part have $\sigma < 0$ and $\omega \neq 0$. Similarly, an unstable spiral would have dominant eigenvalues with $\sigma > 0$ and $\omega \neq 0$, whereas a saddle would have dominant eigenvalue with $\sigma > 0$ and $\omega = 0$. 

 We represent in Fig. \ref{fig:stab-sim}(a) these different types of equilibria in the $(V_2/V_1, V_3/V_2)$ space , for two selected haematocrits and $w=11\,\mu$m.  The stable states correspond to UU and DD configurations, which is in agreement with the experimental observation depicted in Fig. \ref{fig:V_ratios_function_time_and_concs_ALL}(b): for the narrowest channel, if we exclude the states where the flux in the bridges are within the experimental uncertainty established while determining the reference state, the probability to be in UU or DD state is 78\%, to be compared with a 22 \% likelihood to be in a UD or DU configuration. The system indeed seems to spend more time in the  flow configurations that correspond to the stable states predicted by stability analysis.
 Generally speaking, the long transit times through the network (due to the slow flow in the bridges) partly dictates the linear response to perturbations and leads to
growth (decay) rates that are 
very large compared to
the bulk transit time in the network. We therefore expect to find slowly evolving dynamics coupled with fast transitions mediated by the saddle points. 

We confirm these theoretical predictions with independent time-domain simulations of the full nonlinear advection equations, as shown for the perturbed network in Fig. \ref{fig:stab-sim}(b). 
The inlet haematocrit is initially ramped-up from $H_D=0$ to $H_D=0.7$ over the course of $\tau=t/t_0=20$ transit times and the system converges to the stable DD-state. At $\tau=50$ we instantaneously increase the inlet haematocrit to $H_D=0.8$ and the system gets kicked toward a sustained oscillation in the vicinity of an unstable spiral. If we return to $H_D=0.7$ after 29 transit times, the network is attracted to the stable UU-state. In contrast, if we return to $H_D=0.7$ after 33 transit times, the network is attracted to the stable DD-state, \textit{i.e.} the outcome depends on the interplay between the basin of attraction of each stable equilibrium and the state of the system when the inlet haematocrit is changed.

The time domain simulation shown in Fig. \ref{fig:stab-sim}(b) exhibits  transitions between  states that resemble that seen in experiments (Fig. \ref{fig:V_ratios_function_time_and_concs_ALL}(a)) and suggests that the system may converge, after many bulk transit times, to a stable, steady state or a sustained oscillation.

\section{Relevance to physiological flows}

Importantly, the values of haematocrit and channel sizes at which symmetry-breaking occurs in our experiments are 
near the range of human microcirculatory physiological conditions ($H_D\approx 0.5$ and $w\approx 10~\mu$m, corresponding to an average tube haematocrit $H_T \approx 0.37$). The discharge haematocrit $H_D$ is a conserved flux ratio, indicating that microvessels are typically perfused with this haematocrit, with large fluctuations around this average value due to phase separation \cite{baskurt}. This makes this study relevant for a better understanding of RBC distribution heterogeneities in \textit{in vivo} networks.

Note that while we  initially used in our simulations  an \textit{in vitro} viscosity law \cite{Pries94} to tentatively match \textit{in vitro} experiments,  similar computations using \textit{in vivo} data laws show a lower threshold within physiological ranges:  Pries \emph{et al.} have also supplied a widely used quantification of the F\aa hr\ae us--Lindqvist effect \textit{in vivo} \cite{Pries94}:
\begin{align}
\eta_{\mbox{vivo}} &= \left[1 + z(\eta^* - 1) {(1 - H^D)^C - 1 \over (1 - 0.45)^C -1} \right]z \label{eqn:pries_vivo} \\
\eta^* &= 6 e^{-0.085w} + 3.2 - 2.44 e^{-0.06w^{0.645}} \\
z &= \left({w \over w - 1.1}\right)^2,
\end{align}
where $C$ and $y$ are identical to the \textit{in vitro} formulation, and $w$ and $H_D$ are again the the vessel diameter in microns and the discharge haematocrit, respectively. Due to this change in viscosity law, and more precisely to a larger logarithmic derivative \cite{geddes10,karst13},  we note a significantly lower multistability threshold, $H_D \approx 0.41$. 

\begin{figure}    
     \includegraphics[width=0.7\linewidth]{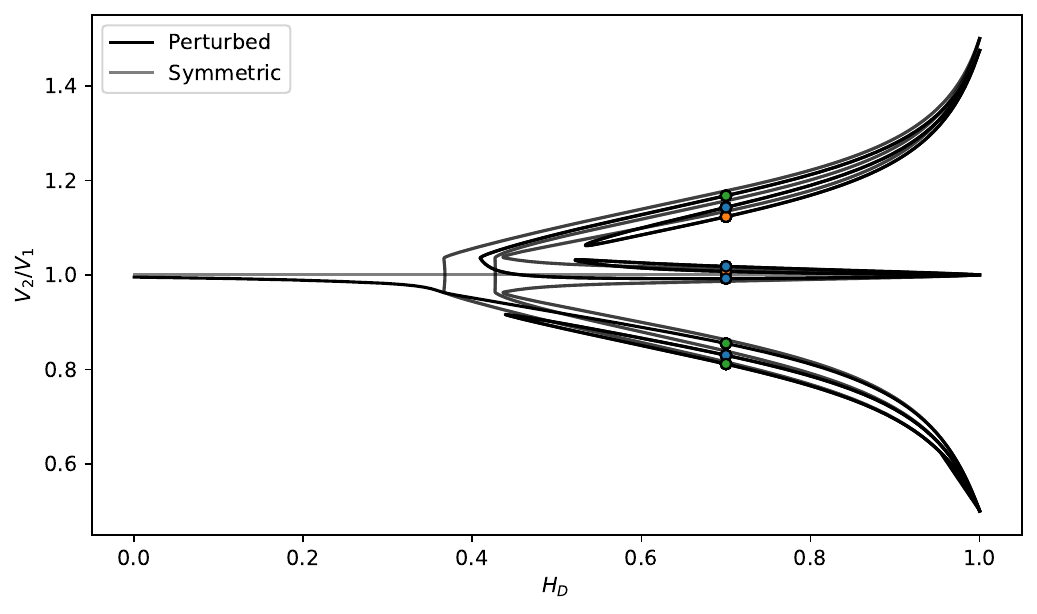}

  \caption{Bifurcation diagram for the symmetric network (semi-transparent marks) and geometrically perturbed network (opaque marks) with $w=11 \, \mu$m and \textit{in vivo} viscosity. 
  \label{fig:bifurcation_diagram_vivo} }
\end{figure}

The existence of a threshold well within the physiological range is consistent with previous studies on complex networks \cite{karst17, karst23}. It also suggests  the existence of lower thresholds \textit{in vivo} compared to that we exhibited in our \textit{in vitro} experiments.

\section{Discussion and conclusion} By experimentally characterizing the evolution of velocity and haematocrit distributions in a simple, nearly symmetric network made of two bridges, we demonstrate that multiple, non-trivial flow solutions emerge and that
haematocrit and channel sizes are the main parameters triggering 
this emergence, with thresholds in a physiologically relevant range.
These multiple solutions are favored by the existence of multiple paths: blood flow in a single loop with no bridge exhibits a single solution \cite{Pop:2007}.  Of critical importance, these solutions 
are predicted by a continuous model based on few ingredients without needing to invoke discrete events as was the case in several previous studies
\cite{cybulski15,Cybulski2019,li2023}. This suggests to use perturbation methods of this continuous model to account for the flow fluctuations in microcirculation.

While in principle the system should converge to a steady state or a sustained oscillation, we have also highlighted that the long-term outcome, and the values of the associated flux, are highly sensitive to both the haematocrit and geometrical imperfections, as exemplified by the difference between the symmetric and geometrically perturbed networks in Fig. \ref{fig:stab-sim}(a), for which we simply introduced a diameter variation of 1\%. These imperfections, or haematocrit fluctuations, can also be dynamically generated by local clogging/adhesion of circulating cells and should be considered as always present, yet fluctuating, in a network. Since at the same time the system always converges slowly due to the control of the dynamics by lateral bridges and their low velocity flux, we therefore expect that such a system will mostly be in a non-steady situation. These findings suggest that the usual assumption of a single stable state would deserve to be revisited, even in rather simple networks, and that functional "disordered states" may emerge even in the absence of pathologies.

In our simplified mesh-like network, the  identified solutions favor RBC flow in the branches that do not belong to the most direct path and help homogenize the distribution of cells, especially when it fluctuates between different configurations. In the meantime, the cells flowing in these transverse paths will spend more time in the network. We therefore expect the distribution of transit times 
to broaden towards large values. The potential analogy with the broadening of distribution  of transit times obtained as multiple paths in space are explored in mesh-like networks \cite{goirand21} deserves to be explored.

\begin{acknowledgments}

MA, GC and TP acknowledge financial support by LabEx Tec 21 (Investissements d'Avenir - grant agreement ANR-11-LABX-0030). GC and TP are grateful to B. de Vicente, G. Lento and M. Leonardo
for preliminary experiments, and to M. van Melle-Gateau for technical support for photolithography. 
TP and GC acknowledge support from  CNES (Centre National
d'Etudes Spatiales, DAR ID 4759).

\end{acknowledgments}

\appendix

\section{Experimental Methods} \label{sec:appendix-exp}

\subsection{Sample preparation}

Blood samples were collected at the \textit{Etablissement Fran\c{c}ais du Sang (French Blood Agency, EFS)}
 and stored in a citrate solution to prevent coagulation.
The blood preparation protocol follows the prescriptions  of \citet{merlo23}. Cells are washed 3 times in Phosphate-Buffered Saline (PBS, P4417 from Sigma Aldrich) by centrifugation, before being mixed with a buffer solution made of 68.5\% PBS, 31.5\% Optiprep (Axis Shield) and 2~g/L Bovine Serum Albumin (BSA, A7906 from Sigma). The haematocrit (volume fraction in RBCs)  $H_{D0}$ of the preparation to be injected in the network is computed from the initial haematocrit $H_i$ in collected tubes provided by  \textit{EFS} and the initial sample volume $V_i$ and final prepared volume $V_0$  through: $H_{D0}=H_i V_i/V_0 $. Samples are used straight after preparation.

\subsection{Experimental set-up}

The chip is fabricated by pouring  and curing polydimethylsiloxane (PDMS) on a mould produced by direct soft lithography (\textit{Klo\'e, Dilase 250} and photoresist \textit{SU8 1070, Gersteltec}). It is then bonded on a glass slide after plasma treatment. All channels in the 2-bridge network have equal length $L=0.5$~mm, equal height $h=30~\mu$m, and equal width $w$ that was varied from $11$ to $36~\mu$m.

Fig. \ref{fig:scheme}(a) shows the 2-bridge network and upstream channels with two distinct inlets fed respectively either by highly concentrated RBCs (haematrocit $H_{D0}\simeq 60-70 \%$), or  buffer solution. They merge in a converging bifurcation located $11$~mm upstream the network ($\approx 300$ times the width of the channel), ensuring an homogeneous mixing of the RBCs with the buffer solution in the $\approx 38\times 30 \,\mu$m cross section serpentine, thanks to both wall-induced lift and shear-induced diffusion \cite{grandchamp13}. The channels between the inlets and this bifurcation have a length of $1.4$ mm, width $38~\mu$m and thickness $h=30~\mu$m. The flow rate in each inlet is varied using a pressure controller \textit{Elveflow OB1} offering a good stability over time in the pressure range $0-2~\textrm{bar}$. 

The first step consists in filling the network with the buffer solution for 2 hours, allowing for adsorption of BSA on the walls of the channels to prevent cell adhesion. Then both inlet are pressurized at the desired values in order to reach the desired velocities and the targeted haematocit $H_D$ to feed the 2-bridge network. These values are typically, for the RBCs inlet and buffer inlet respectively: 30 mbar - 30 mbar for intermediate haematocrit (as in Fig. \ref{fig:scheme}(a)), 45 mbar - 15 mbar to reach $Q_{buffer}=0$ such that $H_D=H_{D0}$ ((as in Fig. \ref{fig:scheme}(b)), 15 mbar - 45 mbar  to reach $Q_{RBCs}=0$ for rinsing procedures. The sum of the inlet pressures is kept constant so as to ensure that the total flow rate stays in a narrow range, therefore the RBC maximal velocities in the network are always in the physiological range 0.5-2 mm.

Fundamental aspects of this set-up are first that different and continuously adjustable haematocrits can be injected in the network without preparing and handling different samples in the reservoirs, thus allowing for an easy control of the fundamental parameter $H_D$. Second, rinsing 
of the network is achieved in a smooth way and in a reasonable time of a few seconds  without increasing hydrodynamic stresses in the network. The ability to easily rinse the system is important since observed flow patterns could be the consequence of undesired adhesion events of cells, or clogging by dusts, which cannot be detected when the network is filled with a highly concentrated suspension. Therefore, after each measurement, rinsing is systematically made by setting $Q_{RBCs}=0$ while keeping the sum of pressures at the inlets constant in order not to create additional stress that would remove the clogging without noticing it. This enables a visual check of possible cell adhesion or clogging in the network. In that case, the whole  data set acquired from the last rinsing to this point is considered as untrustful and discarded.

For both short and long time  measurements, we set  the desired ratio of RBCs to buffer flow rates after rinsing, then wait around $30~s$ before triggering the acquisition. This time is generally enough to reach a quite constant haematocrit at the inlet of the network inlet (see Fig. \ref{fig:V2V1_ft_complement}).

\subsection{Image acquisition} 

The microfluidic chip is set on the stage of an \textit{Olympus IX71} inverted microscope, illuminated with an external light source. The image is magnified 32 times, resulting in a pixel size of $1.08~\mu$m. For each acquisition, i.e. each measurement of velocities in the network, $260$ images are recorded using a \textit{Phantom V2511} camera of 1280$\times$800 pixels resolution, an exposure time $t_{exp}=5~\mu m$ and a sampling frequency $f_{ech}=1000~$Hz. We ensured that the number of images acquired at this frequency is enough to get converged data for the velocity. For long measurements, an \textit{Arduino} microcontroller triggers the camera every $5$ seconds. Due to RAM limitations, this time step is increased to $20$ seconds after $\approx 400$ seconds of acquisition.\\

\subsection{Image processing: haematocrit}

Two types of haematocrit are usually  defined. The \textit{tube haematocrit} corresponds to the instantaneous volume fraction of RBCs averaged over the channel cross-section $A$:
\begin{equation}\label{eqt:HT}
    H_T = \frac{1}{A} \int H dA,
\end{equation}
where $H$ is the local volume fraction. The \textit{discharge haematocrit}, also called \textit{reservoir haematocrit}, is the volume fraction of RBCs weighted by their velocity:
\begin{equation}\label{eqt:HD}
    H_D = \frac{1}{Q} \int u H dA,
\end{equation}
where $u$ is the local velocity and $Q=\int u dA$ the global flow rate. $H_D$ is therefore the relative RBC flux.

In a large channel, where the surrounding fluid and the RBCs flow on average at the same velocity, these two parameters coincide. In narrow channels, RBCs tend to accumulate in the center and therefore flow faster on average than the suspending medium. As a consequence, $H_T<H_D$. This effect is called the F\r{a}hr\ae{}us effect. Contrary to the tube haematocrit, the discharge haematocrit, which is the ratio of the flux of RBCs against the total flux, is a conserved quantity along channel, whatever its section. It is therefore a relevant quantity to be used as a control parameter for comparing 2-bridge networks of different widths $w$.

It is computed in the inlet area upstream the network (see Fig. \ref{fig:scheme}(a)) by measuring light absorption following a Beer-Lambert approach:

The mean tube haematocrit is first computed from grey level intensities measured in the  region of interest (ROI) whose top and bottom boundaries are set as close as possible from the channel walls. The mean intensity $I$ is obtained by averaging in space and time, in the ROI. The optical density is then computed as: 
\begin{equation}\label{eqt:optical density}
    OD = -\log \left[ \frac{I}{\alpha I_r} \right],
\end{equation}
where $I_r$ is the reference averaged intensity  in the same ROI when no RBCs flow in, obtained from a different acquisition. Since several minutes elapsed between the acquisitions from which $I$ and $I_r$ are computed, slight ambient light variations may occur and should be taken into account. Hence the coefficient $\alpha=I_{\textrm{PDMS}}/I_{r,~\textrm{PDMS}}$ in Eq. \ref{eqt:optical density}, which is the ratio of the averaged intensity in the PDMS (far away from the channel) computed from the considered acquisition, to that in the reference acquisition at the same location. Eventually, assuming  a Beer-Lambert law as in  \cite{Merlo2022, sherwood14, roman16,grandchamp13}, the tube haematocrit in the ROI is computed as follows: 
\begin{equation} \label{eqt:haematocrit_from_optical_density}
    H_T =  H_{T0} \times \textrm{OD} /\textrm{OD}_0,
\end{equation}
where $\textrm{OD}_0=$ optical density obtained when non diluted suspension is injected ($Q_{buffer}=0$): the tube haematocrit is then $H_{T0} $, corresponding to the known discharge haematocrit $H_{D0}$.

The conversion between tube and discharge haematocrit can then be achieved using an empirical relation established by Pries  \cite{pries92}:
\begin{equation}\label{eqt:H0T}
    H_{T} = H_{D}(H_{D}+(1-H_{D})X),
\end{equation}
where $X=1+1.7e^{-0.35d}-0.6e^{-0.01d}$, and $d$ is the diameter of a cylindrical channel, measured in microns, for which this law has been proposed initially. It has been shown in \cite{audemar2022} that viscosity laws in rectangular cross section channels, that are also a consequence of cell centering, can be well described by the law also proposed by \citet{pries92} for cylindrical channels, by setting $d=\textrm{min}(w,h)$. Following this idea, we use Eq. \ref{eqt:H0T} with $d=h$.

Finally, $H_{T0}$ is obtained from $H_{D0}$ through Eq. \ref{eqt:H0T}, and $H_D$ is obtained from $H_T$, which was calculated thanks to Eq. \ref{eqt:haematocrit_from_optical_density}, by solving Eq. \ref{eqt:H0T}, which provides 
\begin{equation} \label{eqt:HD_Pries}
    H_{D} = -\frac{X}{2-2X} + \left[  \left(\frac{X}{2-2X}\right)^2 + \frac{H_{T}}{1-X} \right]^{1/2.} .
\end{equation}

\subsection{Image processing: fluid velocity}

\begin{figure}
\resizebox{0.45\linewidth}{!}{
\includegraphics{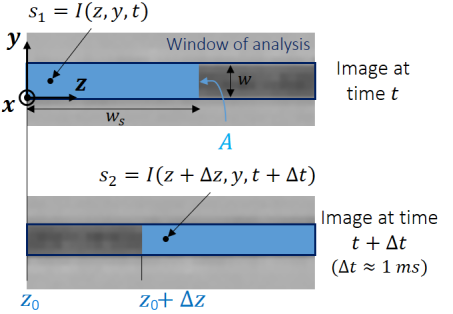}}\caption{Illustration of the method for fluid velocimetry. $\Delta z$ is the distance between the two windows of width $w_s$, and the time step between the two images is of the order of 1 ms.
  \label{fig:Velocimetry_method}}
\end{figure}

The velocimetry consists in calculating an averaged point to point space-correlation of intensities measured at two locations spaced by a varying distance $\Delta z$ along the direction of the flow, separated by a time step $\Delta t$ (see Fig. \ref{fig:Velocimetry_method}). By integrating the space-correlation function over windows of length $w_s$ and width $w$, and considering all the pairs of images available in the recorded sequence, one eventually gets a correlation function whose maximum corresponds to the maximal fluid velocity within the interrogation window. 
 
For a given pair of images $n$, $n+m$ corresponding to times $t=n/f_\textrm{ech}$, $t+\Delta t = (n+m)/f_\textrm{ech}$ ($f_\textrm{ech}$ is the sampling frequency), the measured signals are the grey levels of the pixels located at $(z,y)$ and $(z+\Delta z,y)$ respectively: 
\begin{equation}
\begin{aligned}[b]
& s_1=I(z,y,n),\\
& s_2=I(z+\Delta z,y,n+m)
\end{aligned}
\end{equation}
for all the pixels of the interrogation window  on each image.
The averaged normalized correlation between the two signals is computed on the full range of images and over the areas of analysis:
\begin{equation}
    C(\Delta z) = \frac{\sum_{n,y,z} s_1 s_2}{\sqrt{\sum_{n,y,z} s_1^2} \sqrt{\sum_{n,y,z} s_2^2}},
\end{equation}
where $N$ is the number of images of the acquisition, $\sum_{n,y,z}\equiv\sum_{n=1}^{N-m} \sum_{y=y_0}^{y_0+w} \sum_{z=z_0}^{z_0+w_s}$ denotes the sum over the available pairs of images and over the whole interrogation window, and $(y_0,z_0)$ is the location of the bottom-left corner of the interrogation window on the first image of each pair.

$C_{\Delta z}$ is computed for several values of $\Delta z$, varied from $-20$ to $+20$ pixels, and the fluid velocity $V_{\textrm{f}}$ is deduced from the value of $\Delta z_0$ that corresponds to the position of the  maximum of the correlation function (obtained by a local Gaussian fit to reach subpixel precision): 
\begin{equation}\label{eqt:RBC_velocity_double_slit}
    V_\textrm{f} = \frac{\Delta z_0 \delta}{\Delta t},
\end{equation}
where $\delta$ is the pixel size. As in \cite{roman12} where a time-correlation method was used, we checked that the measured velocity corresponds to the maximal velocity in the $xy$ plane (i.e. central velocity) by applying our method to simulated Poiseuille flows of objects of elliptic cross section with random position and orientations.

In practice, the time interval between the two images must be small enough to get a strong correlation between the signals measured in the two windows, but large enough so that the relative error on the detection of the position of the maximum is small. In practice, this would be reached when the cells are displaced by a distance of the order of a RBC diameter between the two acquisition times, after what correlation is lost because of the complexity of cell dynamics in concentrated suspensions. With $\Delta z_0$ of the order $10~\mu$m, a velocity of the order order $1~\mu$m, $\Delta t$ should be around $10$ ms. Finally, a quick sensitivity analysis has been performed in the range $2 \leq \Delta t \leq 5$ ms for a given dataset: the computed velocity varies within $\pm 5 \%$. Therefore, $\Delta t$ is initially set at 3~ms. If the top of the correlation function cannot be fitted by a Gaussian function (i.e. the correlation function  does not show a clear peak), the process is repeated by increasing $\Delta t$ by steps of 1~ms, and up to $10$ ms. Furthermore, the process is also sensitive to the number of pairs of images processed. We thus studied the convergence of the computed velocity for $N-m=30$, $60$ and $120$ pairs, and observed that $N-m\geq 60$ is enough to get converged data.

\begin{figure}
             \includegraphics[width=\linewidth]{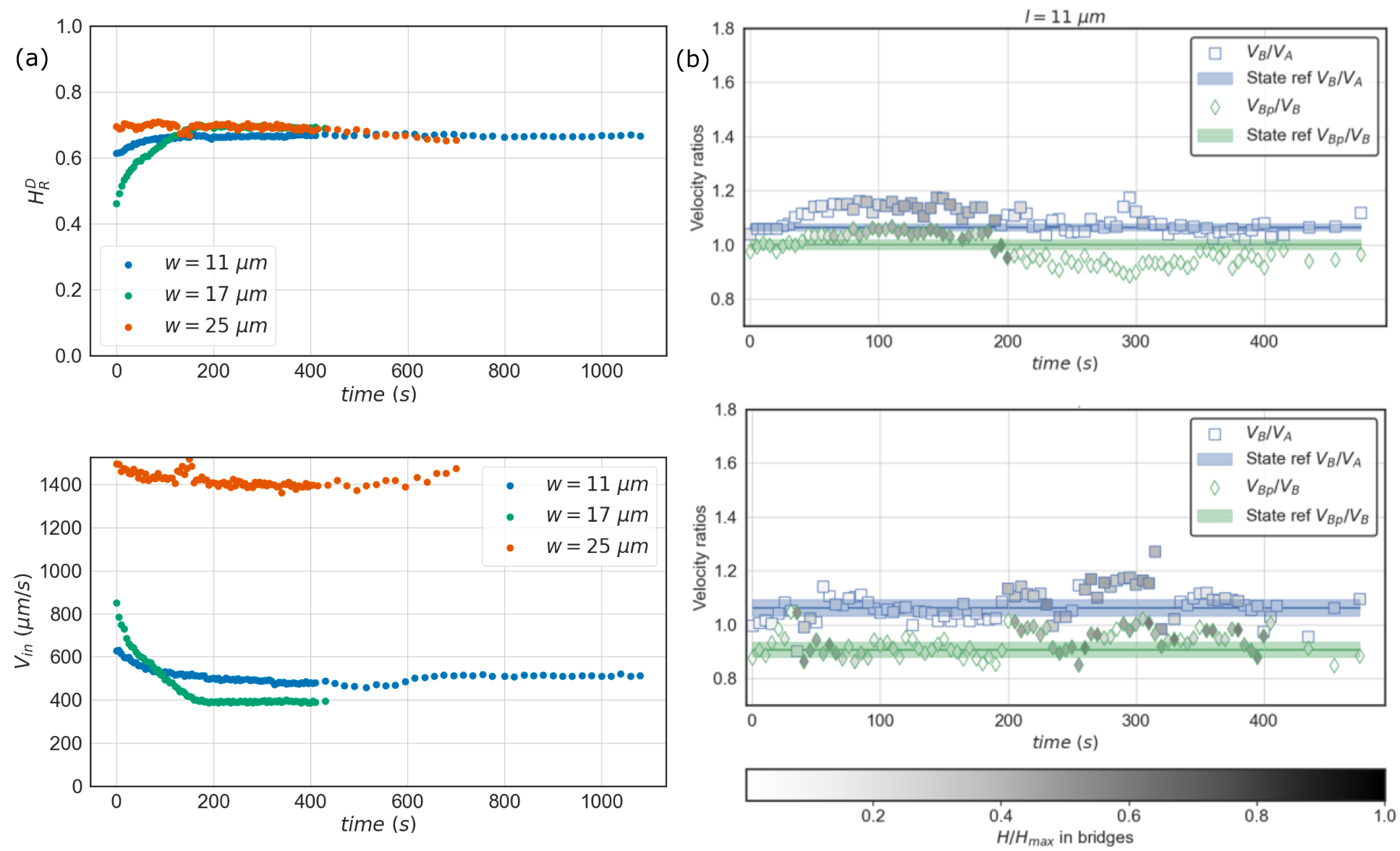}
\caption{(a) Inlet haematocrits and velocities  as a  function of time, for the data shown in Fig. \ref{fig:V_ratios_function_time_and_concs_ALL}(a).(b) Experimental velocity ratios $V_2/V_1$ and $V_3/V_2$ as a function of normalized time, for a channel width $w$ of 11 $\mu$m. Both graphs correspond to two additional experimental realizations with similar parameters as in Fig. \ref{fig:V_ratios_function_time_and_concs_ALL}(a).  The inlet haematocrit $H_D$ is constant in time ($H_D\simeq 0.65 \pm 1\%$). Relative concentrations in each bridge are shown in gray scale normalized by the maximum concentration measured in the bridge. Colored bands identify the reference state, accounting for measurement fluctuations.  \label{fig:V2V1_ft_complement} }
\end{figure}

\subsection{Reference state determination}

In order to compare networks of different widths to build up Fig. \ref{fig:V2V1_f_H}(a), we determine a "reference state" that would be the symmetric state (no flow in the bridges) for a perfect network and is in practice slightly asymmetric because of the imperfections due to the microfabrication process (precision of photolithography, small deformations when bonding the PDMS chip to the glass slide,...).

When $H_D$ was varied, as for the state diagram (short-time measurements) of Fig. \ref{fig:V2V1_f_H}(a),  we simply define the reference velocity ratio as the mean values when $H_D \leq 0.3$, i.e when the fluid is expected to behave as a simple, Newtonian, fluid. However, when processing time series  (Fig. \ref{fig:V_ratios_function_time_and_concs_ALL}), in the absence of low haematocrit data, we first looked for data filling the condition $H_\textrm{bridge}/\textrm{max}(H_\textrm{bridge})<0.05H_{D0}$, i.e. situations with almost no cells in the bridges. If no data could satisfy this condition, which can happen if the bridges are initially filled with cells that then do not move, we then searched  data verifying [$H_\textrm{bridge}/\textrm{max}(H_\textrm{bridge})<0.6H_{D0}$ \& $0.9 \leq V_{n+1}/V_n \leq 1.1$], i.e. situations with small enough flow in the bridges. Taking the mean values for the selected data, we obtain the reference values for the velocity ratios. The colored intervals in Fig. 2 in the main paper correspond to these values $\pm$ the standard deviation. For experiments where low $H_D$ data were available, we found that this method induces a deviation of no more than 8\% compared to the expected value.

\section{Additional data for long time measurements}\label{sec:appendix-ressup}

We plot in Fig. \ref{fig:V2V1_ft_complement} the inlet velocities and haematocrits as a function of time for the experiments leading to the data of Fig. \ref{fig:V_ratios_function_time_and_concs_ALL}(a); these values converge to a quasi steady one in about 100~s. The data shown in Fig. \ref{fig:V_ratios_function_time_and_concs_ALL}(a) are taken when this plateau is reached, i.e  when variations do not exceed 10 \% of the mean value.

Fig. \ref{fig:V2V1_ft_complement} shows two other long-time measurements in the $w=11\,\mu$m network. They complement the graph shown in Fig. \ref{fig:V_ratios_function_time_and_concs_ALL}(a) in the main paper.


%

\end{document}